\documentclass{ws-ijmpa}

\begin{document}

\markboth{Ulf-G. Mei{\ss}ner}
{Challenges in Hadron Physics}

%
\catchline{}{}{}{}{}
%

\title{CHALLENGES IN HADRON PHYSICS}

\author{\footnotesize Ulf-G. Mei{\ss}ner}

\address{Helmholtz-Institut f\"ur Strahlen- und Kernphysik (Theorie), 
Universit\"at Bonn\\
Nu{\ss}allee 14-16, D-53115 Bonn, Germany\\ and\\
Institut f\"ur Kernphysik (Theorie), Forschungszentrum J\"ulich\\
D-52425 J\"ulich, Germany}

\maketitle

\pub{Received (Day Month Year)}{Revised (Day Month Year)}

\begin{abstract}
In this talk, I address some open problems in hadron physics 
and stress their importance for a better understanding of QCD 
in the confinement regime. 

\vspace{-8.9cm}

{\tiny \hfill HISKP-TH-04/16}

\vspace{8.9cm}

\end{abstract}

\section{Introductory remarks}    

Hadron physics explores the least understood sector of the Standard
Model (SM), namely QuantumChromoDynamics (QCD) at large gauge coupling. In the
region of energies below a few GeV, one is faced with two different  
types of challenges:
\begin{romanlist}
\item {\em Precision calculations/measurements:} 
Many very precise data exist which
let one extract fundamental QCD parameters (condensates, quark masses, etc.)
or explore bounds for physics beyond the SM. In such cases, precision
calculations to accuracies of a few percent or better have to be done,
which in certain cases can be achieved utilizing effective field theory
techniques. However, it should be stressed that contrary to popular opinion,
more such precise data are needed, e.g. in processes involving strange
quarks. Space does not allow to discuss these very interesting developments here. 
\item {\em The spectrum:} To really understand the issue of confinement in the
light quark sector, we must first bring order into the spectrum. The nature
and the spectrum of hadrons can only be understood if we are able to say
with certainty which states are genuine quark model states, which are
dynamically generated through channel couplings or which ones are exotic.
For that, precise measurements of many properties of these states are needed
and theoretical tools have to be developed to separate the often overlapping
resonances. In this talk, I will address two topics related to this type of
problems, which are a) some properties of the exotic $\Theta^+ (1540)$ (the
so-called pentaquark) and b) the way to 
achieve a unified description of resonances. Clearly, this choice is 
very subjective, so many other interesting developments and/or challenges could
not be discussed.
\end{romanlist}

\section{Pentaquark issues I: Determining its parity}
The parity of the recently discovered exotic state $\Theta^+ (1540)$
with positive strangeness is not yet determined \cite{PDG} (I refer to
the various plenary talks given at this conference  for a detailed
discussion on the various experiments and possible criticism as well
as on theoretical approaches). 
Its parity is considered
a decisive quantity regarding its substructure. The most appealing 
proposal to determine the parity is based on the application of the
Pauli principle in the process $\vec p \vec p \to \Theta^+ \Sigma^+$
\cite{THK} since it links spin and parity. 
In \cite{CHetal}  the spin correlation coefficient $A_{xx}$ 
was identified as the crucial observable since in the low-energy region
its sign is directly related to parity of the $\Theta^+$, see Fig.~1 and
also Table~1 (assuming that the $\Theta^+$ has spin 1/2). 
Due to the self-analyzing hyperon decay, the energy 
dependence of spin transfer coefficient $D_{xx}$ might also be used 
to determine $\pi (\Theta^+)$ (under certain favorable conditions),
cf. again Table~1 (where the threshold values/ranges are given).
In these arguments, one assumes naturalness of the contributing 
partial waves, in particular for the S-waves that dominate the
region of small excess energy $Q$.

\medskip

\parbox{6.5cm}{
\epsfig{figure=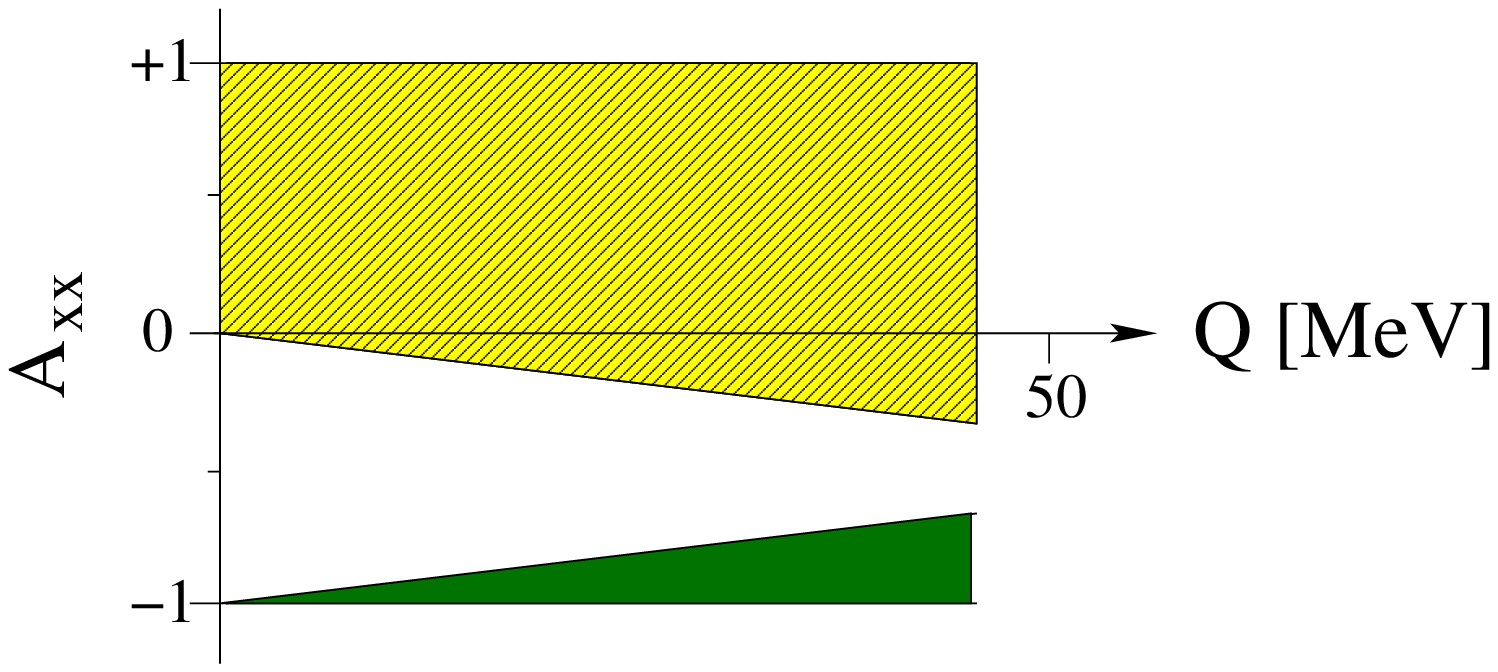,width=6.2cm,height=3.5cm}
\hfill
}
\parbox{5.5cm}{\vspace*{-0.1cm}
{\small \setlength{\baselineskip}{2.6ex}Fig.~1.~Schematic presentation of the
 result for $A_{xx}$  for the two possible parity states of the $\Theta^+$.
For either option realized the corresponding data should fall into the area indicated.
  In case of a negative parity the threshold value depends on the ratio of the
  strength of the two possible S--wave amplitudes.
}}

\begin{table}[h]
\tbl{Spin transfer and spin correlation coefficients for positive/negative
  $\Theta^+$ parity in $\vec p \vec p \to \Theta^+ \Sigma^+$ at threshold. 
For definitions, see e.g. Ref.~\protect\cite{CH}.}
{\begin{tabular}{|cc|cc|cc|cc|} \hline
$D_{xx} (+)$ & $D_{xx} (-)$ & $D_{zz} (+)$ & $D_{zz} (-)$ 
& $A_{xx} (+) $ & $A_{xx} (-)$ & $A_{zz} (+)$ & $A_{zz} (-)$
\\ \hline 
0 & $[-1/\sqrt{2},1/\sqrt{2}]$ & 0 & [0,1]  & $-1$ & $[-1,1]$  &  $-1$ &  [0,1] \\
\hline
\end{tabular}}
\end{table}
\noindent
In \cite{HHKM} are more detailed study of these proposals was performed,
in particular, models were constructed were the leading S-waves are
suppressed due to cancellations, thus explicitely violating the naturalness
assumption. Still, it was shown that the energy dependence of $^3\sigma_\Sigma =
\sigma_0 (1+(A_{xx}+A_{yy})/2)$, with $\sigma_0$ the unpolarized cross section, 
guarantees unambiguous information on the parity of the $\Theta^+$. This can be
achieved e.g. by measurements at two $Q$-values, say at 20 and 40~MeV,
to avoid the Coulomb effects at smaller excess energies but still staying 
sufficiently close to threshold. For a more detailed
discussion (in particular the the usefulness of the quantity $\sigma_0 D_{xx}$)
and more references on this issue, please consult \cite{HHKM}.

\section{Pentaquark issues II: The width of the $\Theta^+$}

Another interesting property of the  $\Theta^+$ is its width. Most experiments
which have seen evidence of this state only quote upper limits for the width
given by the resolution of the experiment. For example, the fairly
clean signal in proton-proton collisions reported by the COSY-TOF
collaboration \cite{TOF} gives $\Gamma_{\Theta^+} \le (18\pm 4)\,{\rm MeV}$
due to the energy resolution.
In various papers, information on kaon--nucleon scattering \cite{Widthpapers}
was analyzed leading to the conclusion that the width of the $\Theta^+$ 
has to be less than a few MeV. A strongly interacting particle with such a
small width could truly be called ``exotic''.
In \cite{SHKM} the impact of the pentaquark
on differential and integrated cross sections for the reaction
$K^+d{\to}K^0pp$, where experimental information is available at kaon momenta
below 640 MeV (the mass range of 1520 to 1555~MeV corresponds to 
kaon momenta in the range $417 \le k_0 \le 476$~MeV), 
was investigated. The calculation utilizes the J\"ulich
kaon-nucleon model (see \cite{JulKN} and references therein) 
and an extension of it that includes the contribution
of a $\Theta^+ (1540)$ with a variable width. The evaluation of the
reaction $K^+d{\to}K^0pp$ takes into account effects due to the Fermi motion
of the nucleons within the deuteron and three--body kinematics. 
 
 
\parbox{7cm}{
\epsfig{figure=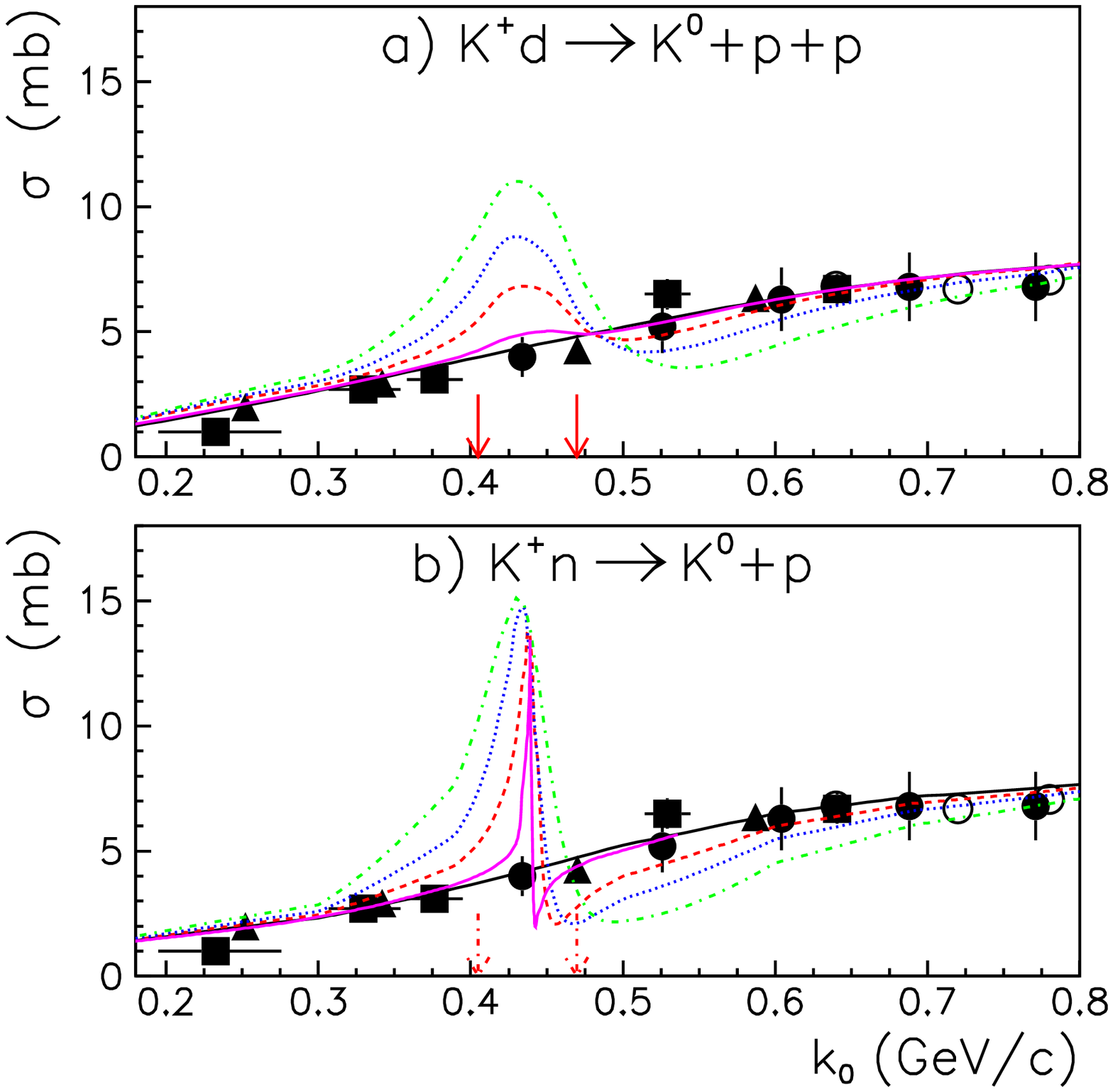,width=6.5cm,height=6.0cm}
\hfill
}
\parbox{5cm}{\vspace*{0.1cm}
{\small \setlength{\baselineskip}{2.6ex}  Fig.~2. 
Total $K^+d{\to}K^0pp$ cross
section as a function of the kaon momentum. 
The lines in a) show our results for the $K^+d{\to}K^0pp$
reaction obtained with different $\Theta^+$  widths:
$\Gamma_\Theta$=1~MeV - solid, 5~MeV -dashed, 10~MeV -dotted
and 20~MeV -dashed-dotted, while the solid (black) line is our calculation
without pentaquark. The lines in b) show the calculations for the
$K^+n{\to}K^0p$ reaction assuming that the neutron target is
at rest.
}}
 
\smallskip\smallskip

\noindent
As shown in Fig.~2, one concludes from that analysis that the data constrain
the width of the $\Theta^+ (1540)$ to be less than 1~MeV. In \cite{SHKM2},
the reaction $K^+ Xe \to K^0 p X$ was investigated in a meson-exchange model
including rescattering of the secondary protons with the aim to analyze the
evidence for the  $\Theta^+ (1540)$ reported by the DIANA 
collaboration \cite{DIANA}. It was confirmed that the kinematical cuts
introduced by the DIANA collaboration efficiently suppress the background to
the $K^+ n \to K^0 p$ reaction which may contribute to the $\Theta^+$
production. These kinematical cuts do not produce a narrow structure  in the 
$K^0 p$ effective mass spectrum near 1540 MeV, cf. the left panel of Fig.~3.
The effect of a narrow resonance with both positive and negative parity
in comparison to the DIANA data was studied in \cite{SHKM2}. It is shown that
the  $K^+ Xe \to K^0 p X$ calculations without $\Theta^+$ contribution as well
as the results obtained with a $\Theta^+$ width of 1~MeV are in comparably
good agreement with the DIANA results, see Fig.~3. The $\chi^2$/dof is 
$2.3$, $2.7$ and $2.9$ for no $\Theta^+$, a positive and a negative
parity spin-1/2 pentaquark with a width of 1~MeV, in order (with the
cuts applied to the $K^0p$ invariant mass spectrum). In view of these
results, the three star rating for the $\Theta^+$ in the recent
PDG tables seems somewhat optimistic \cite{PDG}. More
dedicated experiments are called for to establish (or rule out)
this exotic baryon resonance.

\setcounter{figure}{+2}
\begin{figure}[h]
\begin{center}
\begin{tabular}{cc}
\psfig{file=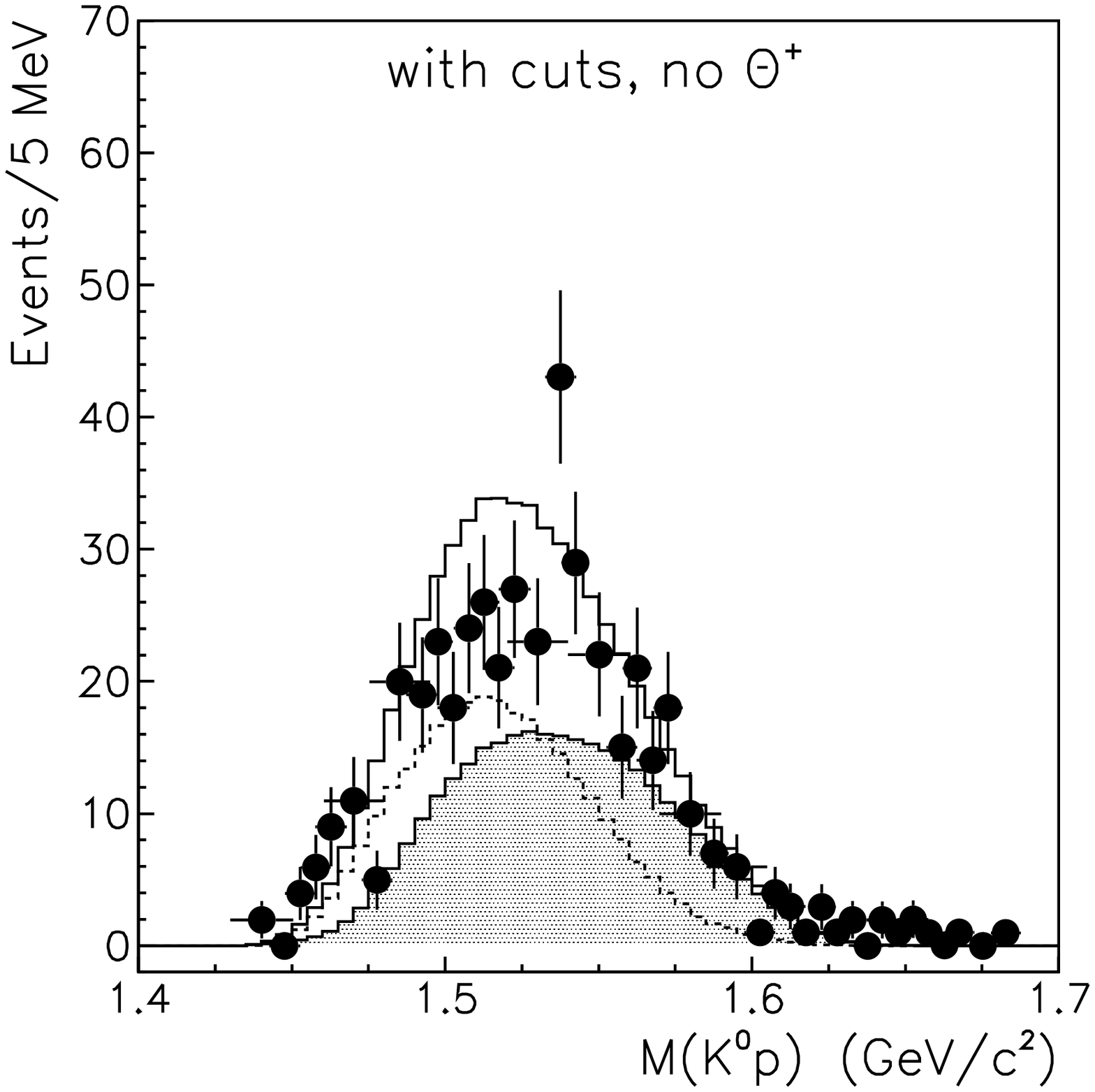,width=5.5cm,height=4.5cm} &
\psfig{file=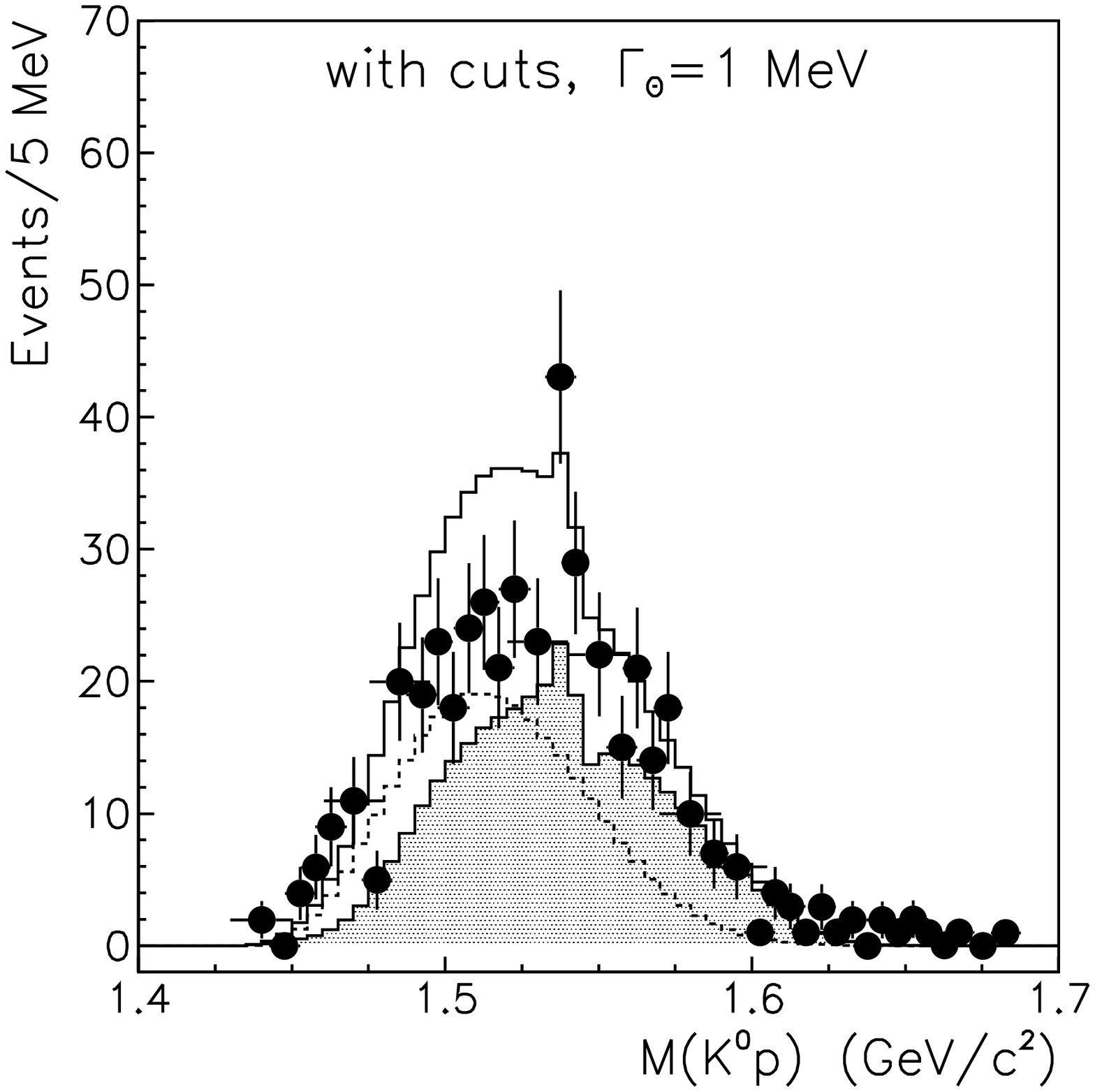,width=5.5cm,height=4.5cm}
\end{tabular}
\caption{The $K^0 p$ invariant mass spectrum from the $K^+Xe$ reaction
with the kinematical cuts performed by the DIANA collaboration. The solid 
histograms are the results without (left)/with the inclusion of a $\Theta^+$
with a width of 1~MeV (right). Hatched histograms show the spectrum  from the
direct $K^+ n \to K^0 p$ production on a bound neutron, for the dashed curve
an additional single rescattering of the proton was taken into account, while
the solid curve is their sum.}
\end{center}
\end{figure}

\section{Unified description of resonances}

This section first presents some results on the baryon spectrum
(for definiteness, a few states are selected to make the point)
using two very different approaches. In the second more speculative
part some ideas are presented how one could unify these approaches
and eventually gain a deeper insight into the confinement regime of QCD.

\subsection{Generating hyperon resonances}

The quark model has been particularly successful in bringing order
into the many observed meson and baryon resonances. In the last years,
it has become evident that a covariant framework is necessary to deal
with {\em all} the (ir)regularities found in the hadron spectrum. 
Here, I will focus on the baryon resonances and pick one particular
covariant quark model, in which the residual interaction of the
constituent quarks are a flavor-dependent instanton induced interaction
and a linear confinement potential (for details, see \cite{LKMP}).
This approach leads to a fairly good description of the spectrum, including
states with very high spin and is consistent with Regge phenomenology.
The 't Hooft interaction also explains the near degeneracy of many positive
and negative parity states \cite{LoeM}, which is very different from the speculations
about chiral restauration in the hadron spectrum at masses above 
$\simeq 1.8\,$GeV which one finds frequently in the current literature.
Many electroweak observables \cite{QMFF} and heavy quark properties have also been
successfully calculated. Strong pion decays come out typically a factor two 
too small, but the overall pattern of pion couplings is well described 
since \cite{MLMP}
the so-called ``missing resonances'' have couplings that are at least an
order of magnitude smaller than for the states that couple with normal
strength to the pion-nucleon continuum. Just to be precise, in this approach
the $\Lambda (1405)$, the  $\Sigma (1620)$ and the $\Lambda (1670)$ are
three--quark states \cite{LMP}. Note, however, that the  $\Lambda (1405)$ is
not well described in this approach, pointing towards the bound-state scenario
discussed below. Furthermore, in the framework of
this model, one also generates
a whole zoo of states with quark content $qqqq\bar q$, some of them exotic
and many of them with conventional quantum numbers, which also has to be
kept in mind in the discussion of the pentaquark $\Theta^+$ - there simply
can not just be one state like that, there must be many (so far one experiment
has a hint for another exotic  baryon from the anti-decuplet \cite{NA49}).
[For a different modern quark model, see e.g. \cite{Graz}].

\smallskip
\noindent
Next, I discuss how hyperon resonances can be generated as meson-baryon
bound states. For that, consider $K^- p$ scattering. A purely perturbative 
treatment is not possible  due to the strong channel couplings and the 
appearance of a subthreshold
resonance, the $\Lambda (1405)$. A  non-perturbative resummation scheme is
mandatory to generate a bound state or a resonance. There
exist many such approaches, but it is possible and mandatory to link 
such a scheme tightly
to the chiral QCD dynamics. Such an improved approach was
developed for pion--nucleon \cite{MeOl1} and later applied to $\bar K$N
scattering~\cite{MeOl2}. 
The starting point is the T--matrix for any partial wave,
which can be represented in closed form if one neglects for the moment
the crossed channel (left-hand) cuts,
$T = 1/[\tilde{T}^{-1} (W) + g(s)]\,$,
with $W = \sqrt{s}$ the cm energy.
$\tilde{T}$ collects all local terms and poles and $g(s)$ is the
meson-baryon loop function (the fundamental bubble) that is resummed by e.g.
dispersion relations in a way to exactly recover the right-hand
(unitarity) cut contributions. The function $g(s)$ needs
regularization, this can be best done in terms of a subtracted
dispersion relation and using dimensional regularization.
It is important to ensure that in the low-energy
region, the so constructed T--matrix agrees with the one of CHPT (matching).
In addition, one has to recover the contributions from the left-hand
cut. This can be achieved by a hierarchy of matching conditions, 
which enforce that to a given order in the chiral expansion, the unitarized
amplitude agrees with the CHPT amplitude (eventually in some unphysical region).
Such a procedure tightly constrains the unitarization procedure, for details
see \cite{MeOl1}. It was observed in \cite{MeOl2} that there
are indeed two poles close to the nominal $\Lambda (1405)$ resonance.
The physics behind these two poles was recently
revealed in \cite{JOORM}. Starting from an SU(3) symmetric Lagrangian to couple
the meson octet to the baryon octet (in that limit, all
octet Goldstone boson masses and all octet baryon masses are equal), 
one could in principle generate a variety
of resonances according to the SU(3) decomposition,
$8 \otimes 8=1\oplus 8_s \oplus 8_a \oplus 10 \oplus \overline{10} \oplus 27~$.
As it turns out, the leading order transition potential is attractive only in the
singlet and the two octet channels, so that one a priori expects a singlet
and two octets of bound states. However, the two octets come out degenerate.
This has no particular dynamical origin but rather is a consequence of
the actual values of the SU(3) structure constants.  In the real world, there
is of course SU(3) breaking of various origins. This was parameterized in
\cite{JOORM} in terms of a symmetry breaking parameter $x$ in the
expressions for  the meson  $M_i$ and baryon masses
$m_i$ as well as the subtraction constants $a_i$ via
$M_i^2 (x) = M_0^2 + x (M_i^2-M_0^2)~,\
m_i (x) = m_0 + x (m_i -m_0)$ and $a_i (x) = a_0 + x (a_i -a_0)$, 
with $M_0 = 368\,$MeV, $m_0 = 1151\,$MeV and $a_0 = -2.148$, where
$0\leq x \leq 1$. The motion of the various poles in the complex 
energy plane as a function of $x$ is
shown in Fig.~4. We note that the two octets split, in particular, one
moves to lower energy ($I=0, 1426\,$MeV) close to the position of the singlet
($I=0, 1390\,$MeV). These are the two poles which combine to give 
the $\Lambda (1405)$ as it appears in various reactions. 
Recent kaon photoproduction data from Spring-8 seem to give 
support to this two-pole scenario.
We note that the $\Sigma (1620)$ and the $\Lambda (1670)$ are also
generated (the latter  is the right-most pole in Fig.~4), 
besides other resonances with strangeness $S = -1$, see
Ref.~\cite{BRO,Valencia} and also the extensive studies in \cite{LK}.

\parbox{7cm}{
\epsfig{figure=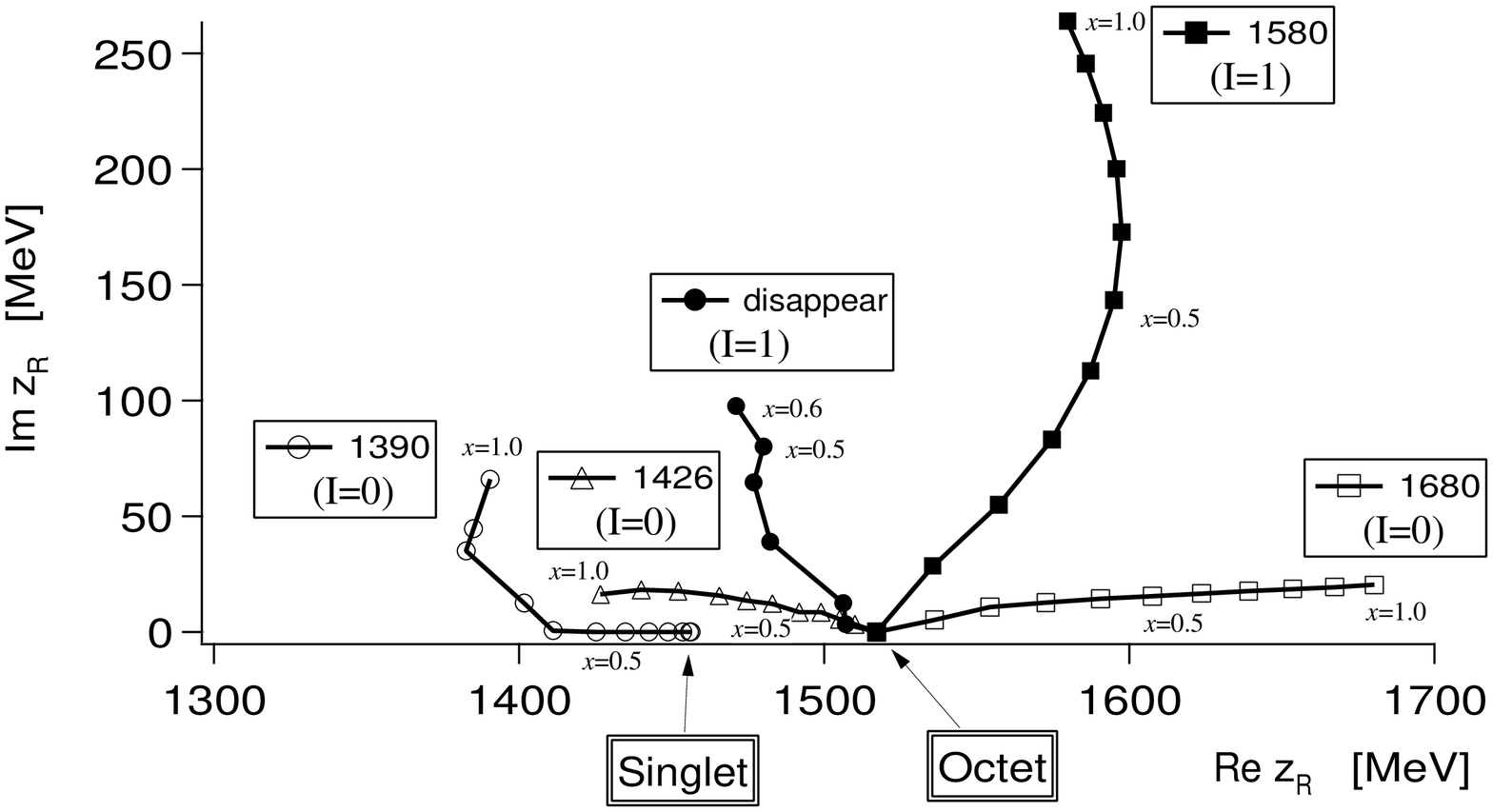,width=6.5cm,height=4.5cm}
\hfill
}
\parbox{5cm}{\vspace*{-0.7cm}
{\small \setlength{\baselineskip}{2.6ex} {Fig.~4.} Trajectories of 
 the poles in
 the scattering amplitudes obtained by changing the SU(3) breaking parameter $x$
 gradually. In the SU(3) limit ($x=0$), only two poles appear, one corresponding 
 to the singlet and the other to the two degenerate octets. 
 The symbols correspond to the step size $\delta x = 0.1$.   
}}

\subsection{Combining the quark model and coupled channel dynamics}

Here, I will speculate how one can bring together the quark model
and the coupled channel approach to achieve a unified description of
resonances. For that, one can follow a two-step procedure, as discussed
below. Before going into the details, I remark that step~1 has been
done (to some extent) whereas the second step so far is science fiction,
but I consider it an important topic to be addressed in the coming years.
More precisely, I envision the following:

\noindent
{\em Step~1:} The coupled channel approach can be extended in that 
one can also include  resonance fields by saturating the
local contact terms in the effective Lagrangian through explicit meson and
baryon resonances (for details, see \cite{MeOl1}). In particular,
in this framework one can cleanly separate genuine quark resonances
from dynamically generated resonance--like states. The former require
the inclusion of an explicit field in the underlying Lagrangian, whereas
in the latter case the fit will arrange itself so that the couplings to
such an explicit field will vanish. This framework was applied in \cite{MeOl1}
to the case of pion--nucleon scattering below the $\eta N$ threshold.
Some typical results are: 1) The delta
pole is located at $(1210, i53)$ MeV, in agreement with dispersion-theoretical
studies of pion--nucleon scattering and pion photoproduction. Furthermore,
a genuine 3-quark component is needed to build up this state. 2) The rho
requires a quark-antiquark component, consistent with large N$_C$
investigations and its tensor coupling comes out large, $\kappa_\rho \simeq
6.3$. 3) No light scalar is needed to describe the data, all the interaction
in the t-channel with scalar-isoscalar quantum numbers is build up by pion
rescattering (loops). This is consistent with the analysis of other reactions,
like e.g. $J/\Psi \to V\pi\pi/\bar KK$ (with $V = \omega, \phi)$ \cite{MOJPsi}
or the scalar
form factor of the pion \cite{UGMscalar}
as well as the intermediate range attraction in the
central part of the two--nucleon potential.

\smallskip\noindent
{\em Step~2:} A further step would be to merge the coupled channel dynamics
with quark model predictions. In Fig.~5 the underlying idea is sketched for
the example of the strong $\Delta \to N\pi$ transition.
In a pure quark model description based e.g. on the Bethe-Salpeter formalism,
one can only describe part of the $\Delta \to N\pi$ decay, since there is
a sizeable modification of the vertex through final state interactions (often
called ``dressing'') and also channel coupling (some typical graphs are
shown on the right-hand-side of that figure). In fact, the idea would be
to use the vertices and masses generated in the quark model as input in these
loop corrections. That way, the aforementioned problem of the too small
strong width should be solved and one also obtains a better representation of
the underlying dynamics. Ultimately, this should be extended to many channels
so as to achieve a truly unified description of the spectrum of the strongly
interacting particles and their properties. To come back to the hyperon states
discussed before, such a scheme would allow one to really pin down the nature
of these states, my guess is that while the $\Lambda (1670)$ and the $\Sigma
(1620)$ are dominantly three-quark states, the  $\Lambda (1405)$ will not
have a three-quark component but simply be a dynamically generated state with
different response to external probes and decay patterns. To end this
section, I should point out that such ideas are not entirely new, but
in my opinion only now the theoretical tools have become available to
seriously tackle this problem.

\setcounter{figure}{+4}
\begin{figure}[h]
\centerline{\psfig{file=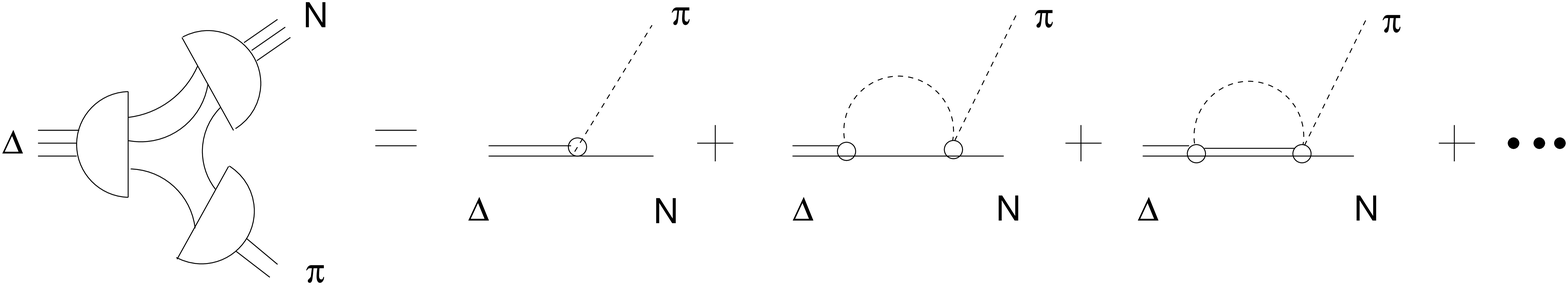,width=9.9cm}}
\vspace*{8pt}
\caption{A schematic representation of merging the quark model with
the coupled channel dynamics. The quark model can be used to generate
the bare particle masses and vertices, which are renormalized in
a consistent fashion (unitarized CHPT) as shown by some typical
one-loop graphs. Higher loop graphs that are not shown here are also 
generated by the unitarization procedure.}
\end{figure}

\section{Concluding remarks}
I would like to stress again that hadron structure and dynamics is one of 
the central issues of QCD (of the Standard Model) and require non--perturbative
methods in the few GeV region and below. Apart from the topics discussed
before, there are many further challenges related to (I just name a few):
\begin{itemize}
\item Lattice QCD and chiral extrapolations. Most lattice data are 
obtained for large unphysical values of the quark (pion) masses and thus
must be interpolated by means of chiral perturbation theory to the
physical value of $M_\pi$. On the example of the nucleon mass, the
possible loopholes and limitations of such methods are discussed in
\cite{BHMlat}. It is evident that for pion masses larger than 500~MeV,
chiral extrapolations cease to be useful.
\setcounter{figure}{+5}
\begin{figure}[h]
\centerline{\psfig{file=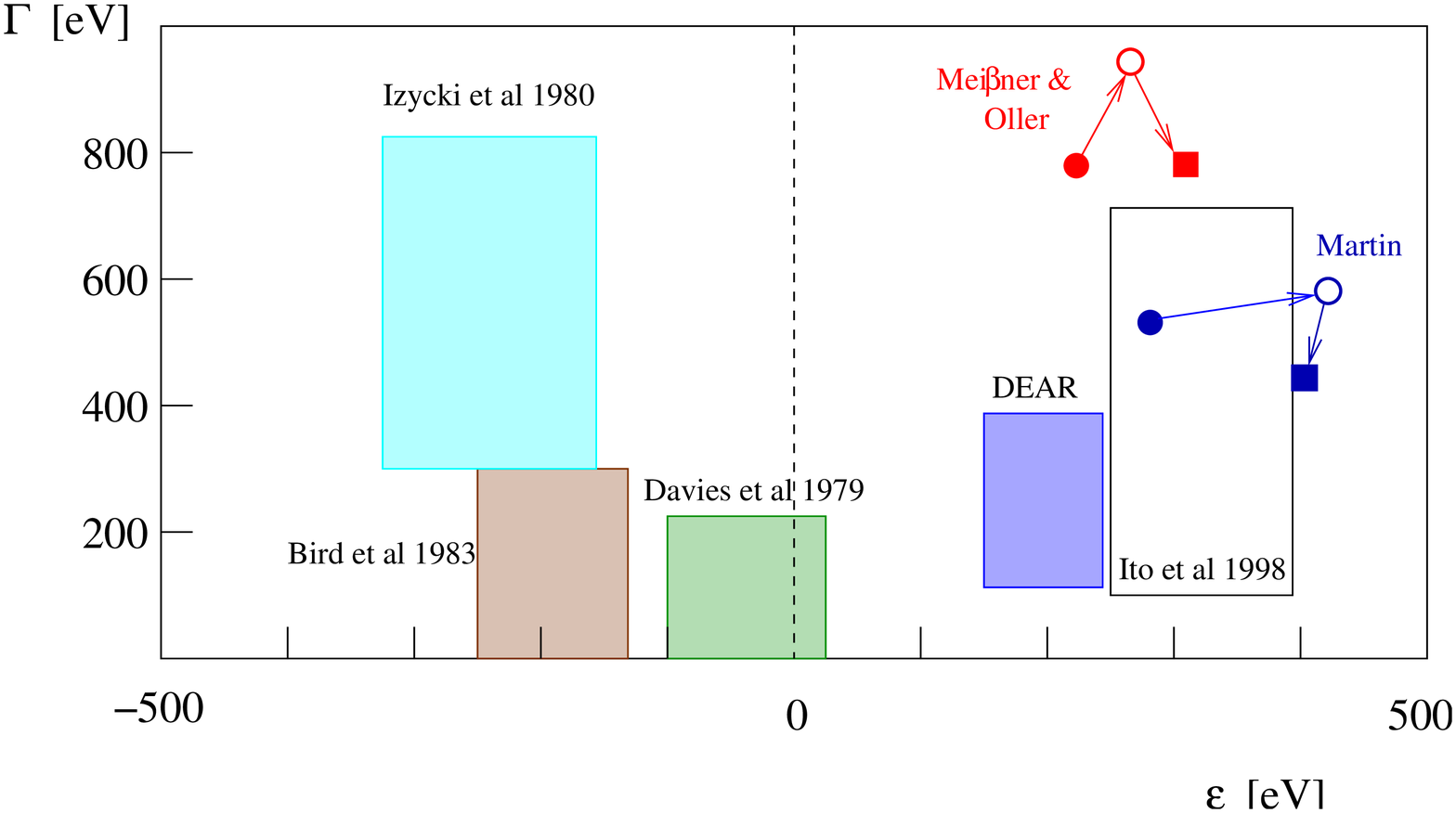,width=9.9cm}}
\vspace*{8pt}
\caption{Predictions for the ground-state strong shift $\Delta E_1^s$ and
the width $\Gamma_1$. Filled circles correspond to using the Deser formula,
empty circles using the scattering amplitude instead of the scattering length (thus
including the large cusp correction due ot the $\bar K^0 n$ channel) and
filled boxes the final formula derived in \protect\cite{MRR}.}
\end{figure}
\item As noted in the introduction, precision hadronic calculations are
needed to e.g.  really be able to address the issue whether physics beyond the
SM has been seen in the precise BNL measurements of the muon anomalous
magnetic moment \cite{BNL} (my guess would be not) or if one wants to precisely test
CKM unitarity in B-decays. Another recent example are the radiative
corrections to tritium $\beta$-decay \cite{GBM} to reduce the upper limit on the
(electron) neutrino mass \cite{Katrin}. 
In addition, there are some real puzzles
when comparing precision data to precision calculations. As an example,
in Fig.~6 a calculation of the ground-state energy shift and width of 
kaonic hydrogen \cite{MRR}
compared to earlier measurements \cite{Kpearlier} and the recent result 
from DEAR \cite{DEAR} is shown. As the only input, 
the S-wave scattering length combination
$(a_0+a_1)/2$ was taken from the dispersive analysis of Martin \cite{Martin}
and from the unitarized CHPT analysis of \cite{MeOl2}. In both cases, the
theoretical prediction (filled boxes) is in stark disagreement with the DEAR
result. At present, this discrepancy is unexplained.
\item There has been impressive progress made in Nuclear Effective Field
Theory. In particular, two- and three-nucleon forces can be described in
one consistent scheme. By now, the nucleon--nucleon force has been worked
out to next-to-next-to-next-to-leading order in the EFT expansion
\cite{EM,EGM} and has reached
the accuracy needed to do precise calculations in few-nucleon systems. 
Other advances in this field concern the quark mass dependence of 
light nuclei \cite{BS,EGMcl} and the speculation of an infrared
renormalization group  limit cycle
in  QCD \cite{BH}. For many other applications, in particular low energy 
reactions of relevance for astrophysics and element synthesis, see e.g.
Refs.\cite{INTrev,Korearev}.
\end{itemize}

\section*{Acknowledgments}

I would like to thank the organizers to give me this opportunity, my
collaborators J.~Haidenbauer, C.~Hanhart, D.~Jido, J.A.~Oller, E.~Oset,
U.~Raha, A.~Ramos, A.~Rusetsky and A.~Sibirtsev 
as well as B.C.~Metsch and H.R.~Petry for sharing
their insight into the topics discussed here. I have also profited from
collaborations with V.~Bernard, E.~Epelbaum, S.~Gardner, H.-W.~Hammer and
W.~Gl\"ockle. Special thanks to Paulo Bedaque and the Nuclear Theory Group
at LBNL for their hospitality where part of this work was done.

\end{document}